\documentclass[runningheads]{llncs}
\usepackage{graphicx}
\usepackage{amsmath,amsfonts}
\usepackage{comment}
\usepackage{hyperref}
\usepackage{array}
\usepackage{caption}
\usepackage{subcaption}
\usepackage{graphicx}
\usepackage{xcolor,colortbl}
\usepackage{romannum}
\usepackage{enumerate}
\usepackage{ragged2e}
\usepackage{multirow}
\usepackage{tabularx,tabulary}
\usepackage{nicefrac}		
\usepackage[flushleft]{threeparttable}
\usepackage{hyperref}
\usepackage{rotating}
\usepackage{tabularx,ragged2e,booktabs}
\usepackage[justification=centering]{caption}
\usepackage{listings}
\usepackage{amssymb}
\usepackage{pifont}
\usepackage{mathtools}
\usepackage{soul}
\usepackage{xcolor}

\usepackage{amssymb}
\newcommand{\cmark}{\ding{51}}%
\newcommand{\xmark}{\ding{55}}%
\usepackage{algorithm}
\usepackage{algpseudocode}
\usepackage[]{todonotes}
\usepackage{float}

\usepackage{tikz}
\usetikzlibrary{calc,angles,positioning,intersections,quotes,decorations.markings}
\usepackage{pgfplots}
\pgfplotsset{compat=1.16}
\tikzset{
	font={\fontsize{14pt}{12}\selectfont}}

\begin{document}
\title{AgEncID: Aggregate Encryption Individual Decryption of Key for FPGA Bitstream IP Cores in Cloud}
%
%


\author{}
\institute{}
\author{Mukta Debnath\inst{1} \and
Krishnendu Guha\inst{2} \and
Debasri Saha\inst{3} \and
Susmita Sur-Kolay\inst{1}}

%
\institute{Indian Statistical Institute \\
\email{\{mukta\_t,ssk\}@isical.ac.in}\\
\and
Cork University  \\
\email{kguha@ucc.ie}
\and
University of Calcutta \\
\email{debasri\_cu@yahoo.in}
}




\maketitle              
\pagenumbering{arabic}
\begin{abstract}
Cloud computing platforms are progressively adopting Field Programmable Gate Arrays (FPGAs) to deploy specialized hardware accelerators for specific computational tasks.
However, the security of FPGA-based bitstream for Intellectual Property (IP) cores from unauthorized interception in cloud environments remains a prominent concern. Existing methodologies for  protection of such bitstreams possess several limitations, such as requiring a large number of keys, tying bitstreams to specific FPGAs, and relying on trusted third parties. 
This paper proposes  {\it AgEncID} (Aggregate Encryption and Individual Decryption), a cryptosystem based on key aggregation to enhance the security of FPGA-based bitstream for IP cores and to  address the pitfalls of previous related works.
In our proposed scheme,  IP providers can encrypt their bitstreams with a single key for a set $S$ of FPGA boards, with which the bitstreams can directly be decrypted on any of the FPGA boards in $S$. 
Aggregate encryption of the key is performed in a way which ensures that the key can solely be obtained onboard through individual decryption employing the board's private key, thus facilitating secure key provisioning. 
The proposed cryptosystem is evaluated mainly on Zynq FPGAs. 
The outcomes demonstrate that our cryptosystem not only outperforms existing techniques with respect to resource, time and energy significantly but also upholds robust security assurances.
\end{abstract}

\section{Introduction}
\label{label:intro}

Field-programmable gate arrays (FPGAs) are increasingly popular in cloud environments due to their reconfigurability. Major cloud providers like IBM, Oracle, Cisco, Microsoft, Amazon, and Google have integrated FPGA reconfigurability into user applications~\cite{data-center}. Reconfiguring FPGAs with different application cores at various times enables efficient task scheduling and execution, resulting in high throughput, low latency, and low power consumption. In public cloud environments, hardware IP cores are obtained from various third-party vendors and deployed on FPGAs connected to the cloud. Instead of users having to acquire these IP cores themselves, they simply transmit their data to the cloud, where it is processed by the application cores within the FPGAs and returned to the users. Safeguarding the confidentiality of these FPGA-based bitstream for IP cores is a critical security concern in this shared cloud environment to prevent unauthorized access and tampering.

\begin{table}[t]
\centering
\caption{Bitstream protection adopted by FPGA Vendors} 
\resizebox{\textwidth}{!}{
\begin{tabular}{|l|ll|ll|l|}
\hline
FPGA vendors:                                                          & \multicolumn{2}{l|}{Xilinx}                                                                                                                                                                                           & \multicolumn{2}{l|}{Altera}                                                                                                          & Microsemi                                                \\ \hline
Devices:                                                               & \multicolumn{1}{l|}{\begin{tabular}[c]{@{}l@{}}Spartan-7,Artix-7,\\ Kintex-7, Virtex-7,\\ Zynq-7000\end{tabular}} & \begin{tabular}[c]{@{}l@{}}Kintex Ultrascale,\\ Virtex Ultrascale,\\ Zynq Ultrascale\end{tabular} & \multicolumn{1}{l|}{Arria 10}                                               & Stratix 10                                             & SmartFusion2                                             \\ \hline
Encryption:                                                            & \multicolumn{1}{l|}{AES-CBC 256}                                                                                  & AES-GCM 256                                                                                       & \multicolumn{1}{l|}{AES-CTR 256}                                            & AES-GCM 256                                            & AES-128/256                                              \\ \hline
\begin{tabular}[c]{@{}l@{}}Key\\ Storage:\end{tabular}                 & \multicolumn{1}{l|}{\begin{tabular}[c]{@{}l@{}}BBRAM/\\ eFUSE\end{tabular}}                                       & \begin{tabular}[c]{@{}l@{}}BBRAM/\\ eFUSE\end{tabular}                                            & \multicolumn{1}{l|}{\begin{tabular}[c]{@{}l@{}}BBRAM/\\ eFUSE\end{tabular}} & \begin{tabular}[c]{@{}l@{}}BBRAM/\\ eFUSE\end{tabular} & \begin{tabular}[c]{@{}l@{}}FLASH\\ SRAM-PUF\end{tabular} \\ \hline
\begin{tabular}[c]{@{}l@{}}Side-channel\\ Protection\end{tabular}      & \multicolumn{1}{l|}{Yes}                                                                                          & Yes                                                                                               & \multicolumn{1}{l|}{Yes}                                                    & Yes                                                    & Yes                                                      \\ \hline
\end{tabular}
}
\label{table:fpgavendor}
\vspace{-0.5cm}
\end{table}

\subsection{Security Attacks on FPGA Bitstream in Cloud}
\label{sec:attacks}
FPGA bitstreams are vulnerable to various types of attacks when moving through the cloud:
\begin{itemize}
    \item An attacker could copy bitstreams through a Man-in-the-Middle (MiM) attack and sell them at a lower price to different cloud service providers. This not only reduces the original provider's profits and market share but also damages their reputation.
    \item An adversary might substitute genuine bitstreams with counterfeit ones through reverse engineering (RE) or insert malicious code such as hardware trojan horses (HTH). This could result in FPGA malfunctions or undesired results~\cite{security_of_cloud}.
    \item The attacker could use side-channel analysis (SCA) to extract confidential information from the bitstream, potentially revealing details about the FPGA's design or the underlying bitstream~\cite{Saha2017}.
    \item If the system software is compromised, it may contain malicious logic or HTH, which could harm the bitstream's performance, durability, or lead to the theft of confidential design data~\cite{sok}.
\end {itemize}

\subsection{Limitations in Existing FPGA Bitstream Protection Techniques}
FPGA bitstream protection is typically achieved through bitstream encryption to prevent unauthorized access, cloning, hardware Trojan insertions, reverse engineering, and tampering attempts.
Modern FPGAs, like those from Xilinx, Intel, and Microsemi, support bitstream encryption using AES as listed in~\autoref{table:fpgavendor}, to defend against these types of passive attacks. 
Relying on vendor-provided symmetric encryption (AES), several studies propose various cryptosystems for encrypting, protecting, and verifying FPGA bitstreams~\cite{plaintext,usenix_hwmetering,compression}.
However, those proposed cryptosystems vary in their approaches to generating and handling cryptographic keys. 
\autoref{table:priorWork} shows some notable prior works in this area, along with where our contribution fits in.  
Furthermore, various alternative methods~\cite{sram,key-obfus,flexible_lic,RTLIP} have been suggested to address the security issues.
We identify four key limitations of these techniques for securing bitstreams in cloud platforms:

\begin{itemize}
  \item \textit{Individual Encryption.} IP providers use FPGA-specific encryption-decryption, requiring developers to encrypt each bitstream for every FPGA, leading to significant key management overhead of time and energy.

  \item \textit{Tied to Specific FPGAs.} Each bitstream is typically associated with a specific FPGA board, limiting the cloud service providers' ability to dynamically allocate resources to meet varying customer demands.

  \item \textit{Third-Party Involvement.} Trusted third parties (TTPs) which play a role in provisioning cryptographic keys to FPGAs, may give rise to unintentional  security vulnerabilities. 

   \item \textit{Resource Overhead.} Existing solutions for key protection based on asymmetric cryptographic involves intricate cryptographic operations on FPGAs or necessitate changes to FPGA structures, often resulting in significant resource demands in terms of hardware, time, and power consumption.
\end{itemize}

\begin{table}[t]
\centering
\caption{Prior Works on FPGA Bitstream Protection}
\resizebox{\textwidth}{!}{
\begin{tabular}{|c|ccc|cccc|c|c|c|c|}
\hline
\multirow{3}{*}{\textbf{Works}} & \multicolumn{3}{c|}{\textbf{Bitstream Encryption}}                                                                                                                    & \multicolumn{4}{c|}{\textbf{Key Encryption}}                                                                                                                                                                 & \multirow{3}{*}{\textbf{\begin{tabular}[c]{@{}c@{}}FPGA\\ Cloud\end{tabular}}} & \multirow{3}{*}{\textbf{\begin{tabular}[c]{@{}c@{}}TTP\\ Required\end{tabular}}} & \multirow{3}{*}{\textbf{\begin{tabular}[c]{@{}c@{}}Change in\\ FPGA\end{tabular}}} & \multirow{3}{*}{\textbf{\begin{tabular}[c]{@{}c@{}}Mapping\\ (IP : FPGA)\end{tabular}}} \\ \cline{2-8}
                                & \multicolumn{1}{c|}{\multirow{2}{*}{\textbf{\begin{tabular}[c]{@{}c@{}}Crypographic\\ Technique\end{tabular}}}} & \multicolumn{2}{c|}{\textbf{Protect Against}}       & \multicolumn{1}{c|}{\multirow{2}{*}{\textbf{\begin{tabular}[c]{@{}c@{}}Crypographic\\ Technique\end{tabular}}}} & \multicolumn{3}{c|}{\textbf{Protect Against}}                                              &                                                                                &                                                                                  &                                                                                                   &                                                                                         \\ \cline{3-4} \cline{6-8}
                                & \multicolumn{1}{c|}{}                                                                                           & \multicolumn{1}{c|}{\textbf{RE}} & \textbf{Cloning} & \multicolumn{1}{c|}{}                                                                                           & \multicolumn{1}{c|}{\textbf{MiM}} & \multicolumn{1}{c|}{\textbf{Tampering}} & \textbf{SCA} &                                                                                &                                                                                  &                                                                                                   &                                                                                         \\ \hline
Guneysu et al.\cite{Guneysu}                  & \multicolumn{1}{c|}{AES-256}                                                                                    & \multicolumn{1}{c|}{\cmark}           & {\cmark}                & \multicolumn{1}{c|}{ECDH}                                                                                       & \multicolumn{1}{c|}{N/A}          & \multicolumn{1}{c|}{\cmark}                  & {\cmark}            & {\xmark}                                                                             & {\cmark}                                                                                & {\cmark}                                                                                                  & 1:1                                                                                     \\ \hline
Drimer et al.\cite{drimer}                   & \multicolumn{1}{c|}{AES-256}                                                                                    & \multicolumn{1}{c|}{\cmark}           & \cmark                & \multicolumn{1}{c|}{ECDH; SHA-1}                                                                                & \multicolumn{1}{c|}{N/A}          & \multicolumn{1}{c|}{\cmark}                  & \cmark            & \xmark                                                                             & \xmark                                                                                & \cmark                                                                                                & N:1                                                                                     \\ \hline
Eguro et al.\cite{fpga-trust}                    & \multicolumn{1}{c|}{AES-256}                                                                                    & \multicolumn{1}{c|}{\cmark}           & \cmark                & \multicolumn{1}{c|}{RSA; SHA-256}                                                                               & \multicolumn{1}{c|}{\xmark}            & \multicolumn{1}{c|}{\cmark}                  & \cmark            & \cmark                                                                              & \cmark                                                                                & \cmark                                                                                                 & 1:1                                                                                     \\ \hline
Maes et al.\cite{sram}                     & \multicolumn{1}{c|}{AES-256}                                                                                    & \multicolumn{1}{c|}{\cmark}           & \cmark                & \multicolumn{1}{c|}{H/W Metering}                                                                               & \multicolumn{1}{c|}{N/A}          & \multicolumn{1}{c|}{\cmark}                  & \cmark            & \xmark                                                                              & \cmark                                                                                & \cmark                                                                                                 & 1:1                                                                                     \\ \hline
Boeui et al.\cite{fasten}                    & \multicolumn{1}{c|}{AES-256}                                                                                    & \multicolumn{1}{c|}{\cmark}           & \cmark                & \multicolumn{1}{c|}{RSA, ECC}                                                                                   & \multicolumn{1}{c|}{\cmark}            & \multicolumn{1}{c|}{\cmark}                  & \cmark            & \cmark                                                                              & \xmark                                                                                & \cmark                                                                                                 & 1:1                                                                                     \\ \hline
Khan et al.\cite{usb-dongle}                     & \multicolumn{1}{c|}{AES-256}                                                                                    & \multicolumn{1}{c|}{\cmark}           & \cmark                & \multicolumn{1}{c|}{HAS}                                                                                        & \multicolumn{1}{c|}{\xmark}            & \multicolumn{1}{c|}{\cmark}                  & \cmark            & \xmark                                                                              & \xmark                                                                                & \xmark                                                                                                 & 1:1                                                                                     \\ \hline
Turan et al.\cite{proxy}                    & \multicolumn{1}{c|}{AFGH PRE}                                                                                           & \multicolumn{1}{c|}{\cmark}            & \cmark                & \multicolumn{1}{c|}{H/W Metering}                                                                                           & \multicolumn{1}{c|}{\xmark}             & \multicolumn{1}{c|}{\cmark}                   &        \cmark      &                                \cmark                                                & \cmark                                                                                & \cmark                                                                                                 & 1:N                                                                                     \\ \hline
Bag et al.\cite{multitenantFpga}                      & \multicolumn{1}{c|}{AES-128}                                                                                    & \multicolumn{1}{c|}{\cmark}           & \cmark                & \multicolumn{1}{c|}{KAC\cite{kac}}                                                                                        & \multicolumn{1}{c|}{N/A}          & \multicolumn{1}{c|}{\cmark}                  & \cmark            & \cmark                                                                              & \xmark                                                                                & \cmark                                                                                                 & N:1                                                                                     \\ \hline
\textbf{AgEncID (ours)}           & \multicolumn{1}{c|}{AES-256}                                                                                    & \multicolumn{1}{c|}{\cmark}           & \cmark                & \multicolumn{1}{c|}{Proposed Key Aggregation}                                                                              & \multicolumn{1}{c|}{\pmb{\cmark}}   & \multicolumn{1}{c|}{\cmark}                  & \cmark            & \cmark                                                                              & {\pmb{\xmark}}                                                                                & {\pmb{\xmark}}                                                                                        & \textbf{1:N}                                                                            \\ \hline
\end{tabular}
}
\label{table:priorWork}
\vspace{-0.5cm}
\end{table}

\subsection{Our Contributions} 
To address the four limitations listed above, we presents \textit{AgEncID} (Aggregate Encryption and Individual Decryption), a cryptosystem centered around key aggregation designed to safeguard FPGA bitstreams.
\textit{AgEncID} offers flexible key management and secure provisioning of cryptographic keys.
Below, we outline the key features of \textit{AgEncID}:

\begin{itemize}
    \item \textit{Single-Key Encryption.} To overcome the first limitation, \textit{AgEncID} allows IP developers to encrypt their bitstream with a single key, which can be used to decrypt the bitstream on a set of FPGA boards for execution.
    This significantly reduces the time and energy cost of the system.

    \item \textit{Decoupling from Specific FPGAs.} IPs are not tied to specific FPGAs, enabling cloud service providers to maximize resource utilization and flexibility.
    This principle helps us to avoid second limitation.

    \item \textit{No External TTP.} To circumvent the third limitation, \textit{AgEncID} eliminates the need for a TTP---the bitstream protection key is known only to the bitstream owner. The cryptographic components for key establishment is performed at the FPGA vendor side who is usually considered to be trusted~\cite{fasten,multitenantFpga}. 
    
    \item \textit{Lower Overhead.} \textit{AgEncID} does not require changes to FPGA fabric or a dedicated cryptographic processor on the FPGA, resulting in lower resource overhead. Significant reduction in time, and power requirement is also achieved compared to existing works.
    Therefore, \textit{AgEncID} effectively tackles the fourth limitation as well.
\end{itemize}

The rest of the paper is organized as follows. 
Section~\ref{sec:system-model} presents system and threat model for FPGA-based cloud system. 
Section~\ref{sec:framework} presents our proposed $AgEncID$ along with its correctness especially in terms of key aggregation.
Section~\ref{sec:AgEncID-in-cloud} discusses the application of \textit{AgEncID} in FPGA-based cloud environment.
Its security and performance analysis appear in Section~\ref{sec:analysis}. Experimental results are discussed in Section~\ref{sec:experiments} with concluding remarks in Section~\ref{sec:conclude}.
\section{FPGA-based Cloud System}
\label{sec:system-model}
This section describes the system and threat models pertinent to an FPGA-based cloud system.

\begin{figure}[t]   
    \centering
    \resizebox{0.7\textwidth}{!}{\includegraphics[]{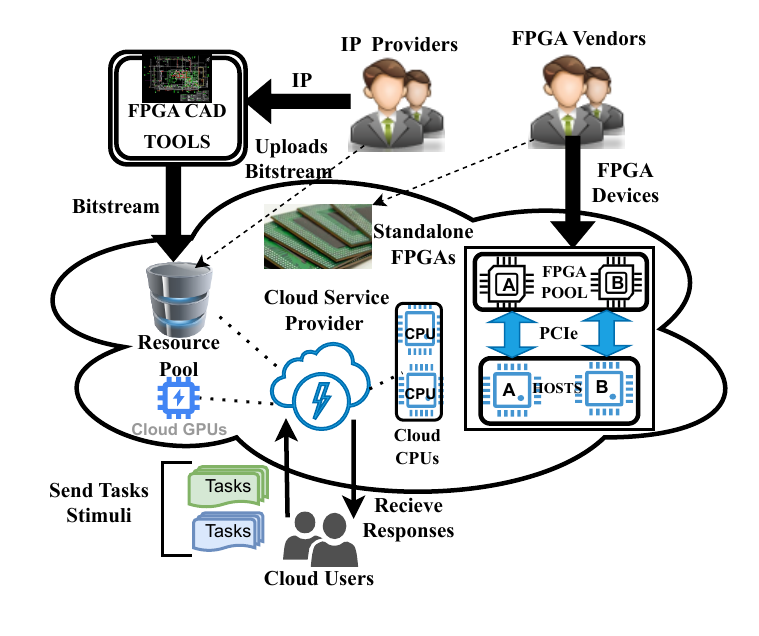}}
    \caption{FPGA-based cloud system model}
    \label{fig:system_model}
\end{figure}

\subsection{System Model}
In an FPGA-based cloud system, there are several key participants involved: FPGA vendors (FVs), IP providers  (IPPs), the Cloud Service Provider (CSP), and the cloud users (CUs) as shown in~\autoref{fig:system_model}. The CSP is responsible for deploying FPGAs from one or more FPGA vendors and offers them as a cloud service to users \cite{fpgatocloud}. To run applications on these cloud-connected FPGAs, the CSP obtains IP cores from various third-party IP providers \cite{trust-in-fpga}. These IP providers provide their designs in the form of HDL, netlists or bitstreams to the cloud.
Cloud users engage with this system by submitting their tasks to the cloud, where application cores within bitstreams run on FPGAs, providing Application-as-a-Service functionality. Multiple users' workloads can share the same physical resources in the cloud, depending on the CSP's allocation and scheduling policies. Typically, commercial CSPs allocate an entire FPGA instance from a cloud-based pool to a single user \cite{trusted_fpga_config}, a model similar to Amazon EC2 F1 (\url{www.aws.amazon.com/ec2/instance-types/f1}). In this setup, the user gets dedicated allotment to the FPGA for a specified period according to their demand and agreement with the cloud provider. However, they do not have direct physical access to the FPGA hardware.

\subsection{Threat Model}
\label{sec:threat-model}
Given the various attacks at play (mentioned in Section~\ref{sec:attacks}), an individual with malicious intentions has the potential to introduce multiple threats to the integrity of the bitstream within the FPGA-based cloud system (Figure~\ref{fig:system_model}).
We considers the following threat scenarios on bitstream.  

\begin{itemize}
    \item \textit{Malicious Cloud Service Provider.} CSPs are not always trustworthy. CSPs can access the RTL design of IP cores and engage in intellectual property (IP) theft by bitstream reverse engineering.
    
    \item \textit{Malicious Cloud User.} Any potential adversaries among the other cloud users who are not authorized for the bitstream usage can steal the bitstream and misuse it without any administrative privileges.
    
    \item \textit{Malicious IP Providers.} The IPPs have access to privileged information about the cloud system and may make efforts to intercept IP bitstreams from other providers when their IPs are running on the same group of FPGAs.
    
    \item \textit{Malicious External Agents.} An external agent may attempts to gain unauthorized access to a bitstream in transit within the cloud.
\end{itemize}
\section{Aggregate Encryption Individual Decryption (\emph{AgEncID})}
\label{sec:framework}
We introduce \textit{AgEncID} (\textit{Ag}gregate \textit{E}ncryption and \textit{I}ndividual \textit{D}ecryption), a cryptosystem based on key aggregation to enhance the security of FPGA-based bitstream IP cores.
\textit{AgEncID} comprises two modules: bitstream encryption module and key aggregation module. 
The bitstream encryption module effectively leverages vendor-provided symmetric encryption, particularly AES, which is widely recognized and does not necessitate extensive elaboration.
Our primary focus lies on the key aggregation module of \textit{AgEncID}, where we introduce a key aggregation technique aimed at enhancing key management efficiency within cloud environments and ensuring the secure provisioning of keys within cloud FPGAs. 
In the subsequent section, we embark on a comprehensive exploration of this proposed key aggregation method implemented within the \textit{AgEncID} cryptosystem.

\subsection{Key Aggregation in \textit{AgEncID}}
\label{subsec:AgEncID}
The proposed key aggregation method of \textit{AgEncID} relies on the utilization of bilinear pairing on elliptic curves.
A \textit{bilinear pairing} is a bilinear map defined over elliptic curve subgroups \cite{pairingecc}. This concept is used in previous works on broadcast encryption (BE) \cite{collusion} and key aggregate cryptosystem (KAC) \cite{kac}.

Let $G$ and $G_T$ be two such (multiplicative) cyclic subgroups of prime order $p$ and
$\widehat{e} : G \times G \rightarrow G_T$ be a map with the following properties:
\begin{itemize}
    \item[] \textit{Property 1.} (\textit{bilinear}) For every $g_1 , g_2, g_3 \in G$, we have \\ (\textit{i}) $\widehat{e}(g_1 + g_2, g_3) = \widehat{e}(g_1, g_3) \cdot \widehat{e}(g_2, g_3)$; (\textit{ii}) 
            $\widehat{e}(g_1, g_2 + g_3) = \widehat{e}(g_1, g_2) \cdot \widehat{e}(g_1, g_3)$.
    \item[] \textit{Property 2.} (\textit{efficiently computable}) The map $\widehat{e}$ can be computed efficiently.
    \item[] \textit{Property 3.} (\textit{identity}) For every $g \in G$, we   have $\widehat{e}(g, g)=1$.
    \item[] \textit{Property 4.} (\textit{non-degenerate}) For all $g \in G$, we  have $e(g, h) = 1$ for all points $h$ if and only if $h = \infty$.
    \item[] \textit{Property 5.} (\textit{alternation}) $\widehat{e}(g_1, g_2)= \widehat{e}(g_2, g_1)^{-1}$ for every $g_1, g_2 \in G$.
\end{itemize}
\noindent

The basic steps of the proposed key aggregation are inspired from  KAC \cite{kac} based on a public key cryptosystem. 
While KAC supports constant-size ciphertexts and an aggregate key for decryption by a set of users, the proposed key aggregation needs an aggregate key for encryption with individual decryption keys. 
Thus a new aggregate key is generated with corresponding encrypt and decrypt operations. 
The following operations are needed:

\begin{itemize}
    \item[\ding{242}] $\mathbf{Setup(1^\lambda, n)}$
    establishes the public cryptosystem parameters $param$  for the entities, using their respective entity IDs $i$.
    Let $\lambda$ and $n$ be two positive integers, where $n$ denotes the number of entities, and $1^\lambda$ the security level parameter. First, a bilinear group $G$ of prime order $p$ where $2^\lambda \leq p \leq 2^{(\lambda+1)}$, a generator $g \in G$ and $\alpha \in_R Z_p$ are chosen randomly. 
    For $i=\{1,2....,n,n+2,...,2n\}$, $g_i = {\alpha^i}g$ are computed. 
    So, output the system parameter set as $param = \{g, g_1,..., g_n, g_{n+2},....., g_{2n}\}$.
    
    \item[\ding{242}] $\mathbf{KeyGen(}n)$ produces  $PK = \{g, g_1,..., g_n, g_{n+2},....., g_{2n},v\}$ as the master public key, where $\gamma \in Z_p$ is chosen randomly such that the master-secret key $msk=\gamma$ and $v={\gamma}g$.  The $n$ private keys $d_1,....,d_n$ for each entity with IDs $i \in \{1, 2,....., n\}$\ are given by $d_i = {\gamma}g_i = {\alpha^i}v$, where $g_i \in G$,
    and
    $\alpha$ can be safely deleted after $Setup$.

    \item[\ding{242}] $\mathbf{Extract(S)}$ is used to generate a constant size aggregate key for encryption. For a set of entities with their unique IDs $j \in S$, where $S \subseteq n = \{1, 2,.., n\}$ the aggregate key $K_{S}$ is computed as:
    $$K_{S} = \sum\limits_{j\in S} g _ {n+1-j}$$

    \item[\ding{242}] $\mathbf{Encrypt(PK, S, K_{S}, m)}$ is for encryption of a message $m$ (in this case cryptographic key is been referred as message).
    For a plaintext message $m \in G_T$ and set of entities with ID $j \in S $,  $ t \in_R Z_p$ is randomly chosen. The ciphertext is computed for the set using its corresponding aggregate key $K_{S}$. The cipher text  $C = \{c_1,c_2,c_3\}$ comprises of three parts: $c_1=tg$, $c_2=t(v + K_{S})$, and $c_3=m(\widehat{e}(g_n,tg_1))$.

    \item[\ding{242}] $\mathbf{Decrypt(S,i,d_{i},C=(c_1, c_2, c_3))}$ is for decrypting the ciphertext for an entity $i \in S$ using it's unique decryption key $d_{i}$.
    If $i \notin S$, then the output is $null$,
    otherwise, the decrypted message $\widehat{m}$ is obtained as:
    $$\widehat{m}=c_3 \cdot \frac{\widehat{e}(d_i\ + b_{i,S}, c_1)}{\widehat{e}(g_{i},\ c_2)}$$
    \noindent
    where $b_{i,S} = \sum\limits_{(j \in S) \wedge (j \neq i)} g_{n+1-j+i}$.
\end{itemize}

\subsection{Correctness of Key Aggregation in \textit{AgEncID}}
The correctness of the proposed key aggregation approach is founded upon the assurance that decrypting the encrypted message will invariably yield the original message. 
The following two lemmas grounded in the properties of bilinear pairings play a crucial role.

\begin{lemma}
    \label{lemma1}
    In the bilinear group $G$, with elements $g_1$ and $g_2$, and $a, b\in Z$, the following equality holds:
        $\widehat{e}(ag_1, bg_2) = \widehat{e}(g_1, g_2)^{ab}$.
\end{lemma}

\begin{proof}
    Based on the bi-linear \textit{Property 1} as stated in section~\ref{subsec:AgEncID}, we can write the following:
    \begin{align}
        \widehat{e}(ag_1, bg_2) &= \widehat{e}\left(g_1+(a-1) \cdot g_1, bg_2\right) \nonumber\\
        &= \widehat{e}\left(g_1, bg_2\right) \cdot \widehat{e}\left((a-1) \cdot g_1, bg_2\right) \nonumber\\
        & \dots \tag{iterating upto $a$-steps} \nonumber\\
        &= \widehat{e}\left(g_1, bg_2\right)^a \label{eq:lemma11}
    \end{align}
    Iterating in a similar manner with $b$-steps on Eqn.~\ref{eq:lemma11}, we have the following:
    \begin{align}
        \widehat{e}\left(g_1, bg_2\right)^a &= \widehat{e}\left(g_1, g_2\right)^{ab} \nonumber
    \end{align}
    Therefore, we can conclude that,
    \begin{align}
        \widehat{e}(ag_1, bg_2) &= \widehat{e}\left(g_1, g_2\right)^{ab} \nonumber \tag{$\blacksquare$}
    \end{align}
\end{proof}

\begin{lemma}
    \label{lemma2}
    For every $g_1 , g_2, g_3\in G$, the following relationship holds:
    \begin{align*}
        \widehat{e}(g_1 - g_2, g_3) = \frac{\widehat{e}(g_1, g_3)}{\widehat{e}(g_2, g_3)}
    \end{align*}
\end{lemma}

\begin{proof}
    BY {\it Property 1} (bilinear)  in Section~\ref{subsec:AgEncID}, we can express as follows:
    \begin{align}
        \widehat{e}(g_1 - g_2, g_3) &= \widehat{e}(g_1 + (-g_2), g_3) \nonumber\\
        &= \widehat{e}(g_1, g_3) \cdot \widehat{e}(-g_2, g_3) \nonumber\\
        &= \widehat{e}(g_1, g_3) \cdot \widehat{e}(-1 \cdot g_2, g_3) \label{eq:lemma21}
    \end{align}
    
    By applying Lemma~\ref{lemma1} with $a=-1$ and $b=1$, we can rephrase Eqn.~\ref{eq:lemma21} as:
    \begin{align}
        \widehat{e}(g_1, g_3) \cdot \widehat{e}(-1 \cdot g_2, g_3) &= \widehat{e}(g_1, g_3) \cdot \widehat{e}(g_2, g_3)^{-1} \label{eq:lemma22}
    \end{align} 
    
    Therefore, based on equations~\ref{eq:lemma21} and~\ref{eq:lemma22}, we can deduce that:
    \begin{align}
        \widehat{e}(g_1 - g_2, g_3) &= \frac{\widehat{e}(g_1, g_3)}{\widehat{e}(g_2, g_3)} \tag{$\blacksquare$}
    \end{align}
\end{proof}

\begin{theorem}
(Correctness of Key Aggregation in $AgEncID$) The decryption of an encrypted message yields precisely the original message $m$.
\end{theorem}

\begin{proof}
Consider $\widehat{m}$ as the decrypted message corresponding to the original message $m$.
According to operations of proposed key aggregation module mentioned earlier,

\begin{align}
    \widehat{m} &= c_3 \cdot \frac{\widehat{e}\left(\left(d_i + \sum\limits_{(j \in S \wedge j \neq i)}g_{n+1-j+i}\right),c_1\right)}{\widehat{e}(g_{i}, c_{2})} \label{eq:thm1}
\end{align}
By substituting values from key aggregation method's operations, we can express Eqn.~\ref{eq:thm1} as:
\begin{align}
    \widehat{m} &= c_3 \cdot \frac{\widehat{e}\left(\left({\gamma}g_{i} + \sum\limits_{j \in S} g_{n+1-j+i} - g_{n+1}\right),\ tg\right)}{\widehat{e}\left(g_{i}, t\left(pk + \sum\limits_{j \in S} g _ {n+1-j}\right)\right)} \nonumber
\end{align}

\begin{align}
    &=c_3 \cdot \frac{\widehat{e}({\gamma}g_{i} , tg) \widehat{e}\left(\sum\limits_{j \in S}\ g_{n+1-j+i} - g_{n+1}, tg\right)}{\widehat{e}( g_{i}, t{\gamma}g) \widehat{e}\left(g_i, t\sum\limits_{j \in S} g _ {n+1-j}\right)} && \text{(by Property 1)} \nonumber
\end{align}
\begin{align}
    &=c_3 \cdot \frac{\widehat{e}\left(\sum\limits_{j \in S} g_{n+1-j+i}, tg\right)}{\widehat{e}\left(tg, \sum\limits_{j \in S}\ g _ {n+1-j+i}\right) \widehat{e}( g_{n+1}, tg)}  && \text{(using Lemma~\ref{lemma1} and~\ref{lemma2})}\nonumber\\
    &=c_3 \cdot \frac{1}{\widehat{e}(g_{n+1},tg)} && \text{(by Lemma~\ref{lemma2})} \nonumber\\
    &=m \cdot \frac{\widehat{e}(g_{n},tg_1)}{\widehat{e}(g_{n+1},tg)}  && \text{(by Property 1)}\label{eq:thm2}
\end{align}

By applying Diffie-Hellman Exponent assumption~\cite{boneh2005hierarchical} ($\widehat{e}(g_1, g_n)=\widehat{e}(g, g_{n+1})$) on Eqn.~\ref{eq:thm2},  
    $\widehat{m} =m \cdot \frac{\widehat{e}(g_{n},tg_1)}{\widehat{e}(g_{n},tg_1)} 
    = m$.  
This establishes the correctness of the encryption and decryption algorithms of the proposed key aggregation module of $AgEncID$.
\hfill ($\blacksquare$)
\end{proof}
\vspace{-0.5cm}

\begin{figure}[t]
    \centering
    \resizebox{0.6\textwidth}{!}{\includegraphics[]{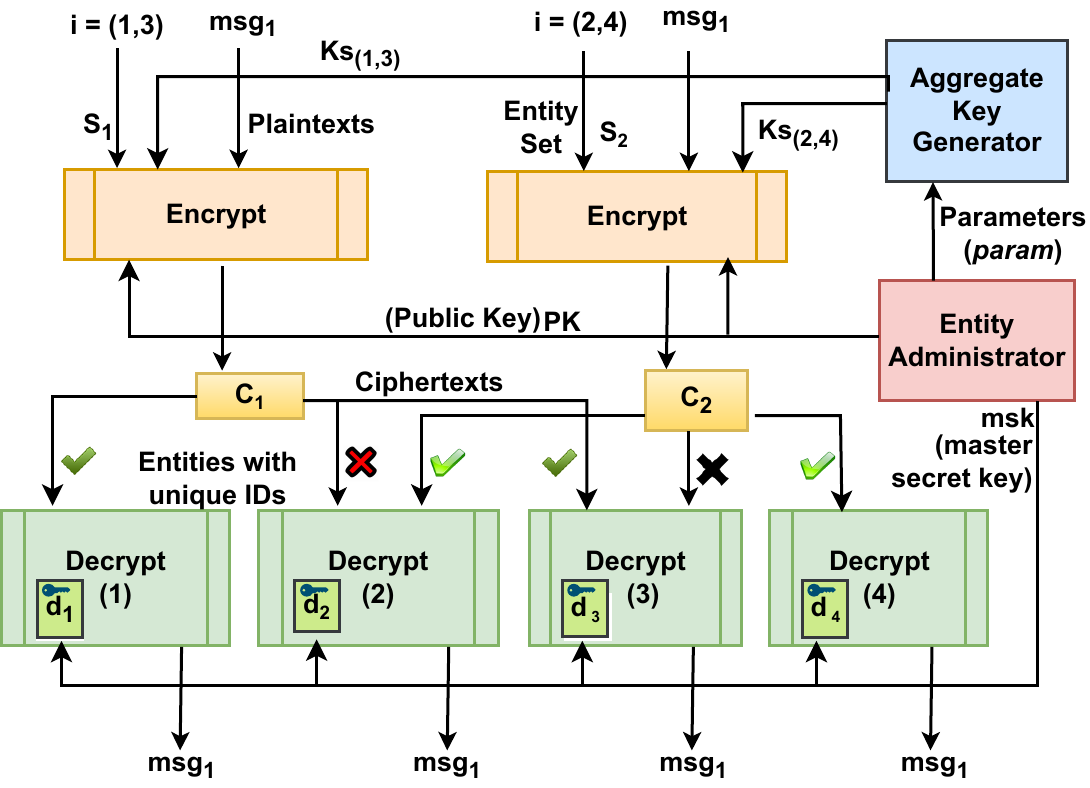}}
    \caption{An overview of key aggregation method of $AgEncID$}
    \label{fig:AgEncID}
\end{figure}

\subsection{A Working Example} 
~\autoref{fig:AgEncID} shows a general illustration of how \textit{AgEncID} cryptosystem works in terms of key aggregation. 
Here, \textit{AgEncID} follows a public-key cryptosystem.
It creates a fixed-size encryption key for a message that needs to be encrypted for various entities, each having its own unique identity ID.
However, it ensures that each entity has its own separate decryption key.
Given a plaintext message, $msg_1$, to be encrypted for a set of $n$ entities with IDs $i=1, 2,.., n$,  it is possible to generate a constant-size aggregate encryption key $Ks$, for any arbitrary subset $S$ of the IDs, $S \subseteq \{1, 2,.., n\}$. The ciphertext can be decrypted by an entity with ID $j$ such that $j \in S$ using its private individual decryption key $d_j$. \autoref{fig:AgEncID} illustrates the $AgEncID$ scheme with two sets of entities represented by their unique IDs in $S_1=\{1,3\}$ and $S_2=\{2,4\}$. 
The entity administrator generates the master public key $PK$ and aggregate keys $Ks_{1,3}$ (for$S_1$) and $Ks_{2,4}$ (for $S_2$). These keys are then employed to encrypt the plaintext message $msg_1$ for both $S_1$ and $S_2$.
The entity administrator also generates the master secret key $msk$, which is used to construct the individual decryption keys. The ciphertext $C_1$ for the message $msg_1$ can be decrypted by entities from $S_1$  using their individual decryption key $d_1$ and $d_3$ but fails to be decrypted by entities from $S_2$ or any other entity set. 

\section{\textit{AgEncID} in FPGA Cloud Environment}
\label{sec:AgEncID-in-cloud}
\begin{figure}[t]
    \centering
    \resizebox{0.9\textwidth}{!}{\includegraphics[]{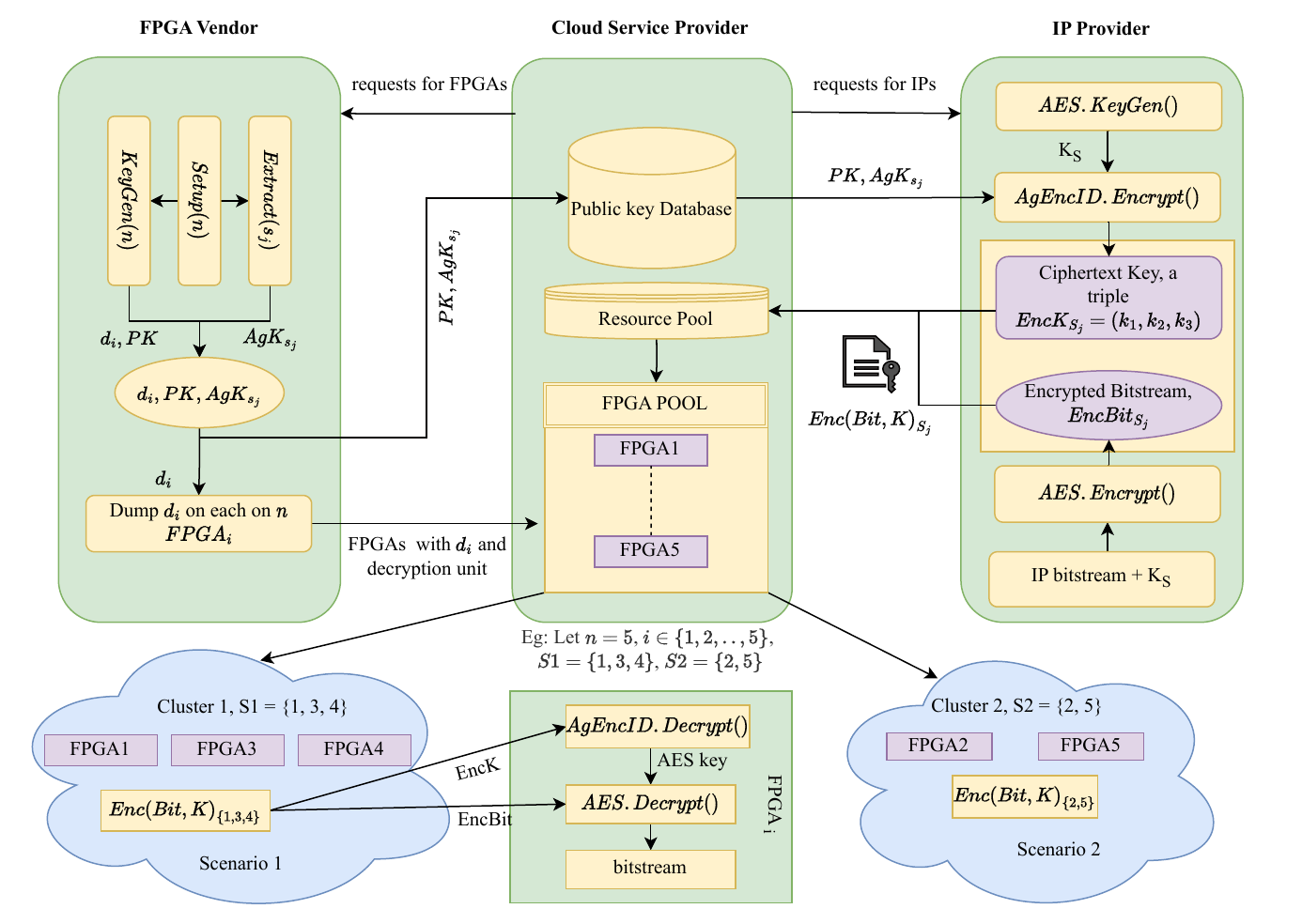}}
    \caption{An example of \textit{AgEncID}'s working principle with one FV and one IPP in FPGA Cloud Environment}
    \label{fig:AgEncID-eg}
    \vspace{-0.5cm}
\end{figure}

This section explains how \emph{AgEncID} works in an FPGA-cloud environment using Algorithm \ref{algo:AgEncID}. 
While a cloud system can have multiple FPGA vendors (FVs), each providing a number of FPGA boards, and IP providers (IPPs), each providing one or more IP cores, here we focus on the communication among a CSP, an  FV and an IPP using $AgEncID$ and provide an example (\autoref{fig:AgEncID-eg}). 
In order to achieve a resemblance with~\autoref{fig:AgEncID}, we can align FPGA boards and FPGA vendors with entities and the entity administrators depicted in the figure.

\medskip

\noindent
\textbf{Cluster Formation for FPGAs.} Cluster Formation for FPGAs involves the CSP grouping FPGA board requirements based on FVs and board families. We consider three scenarios: \\\textit{Scenario 1:} A cluster with $n$ boards from the same FV and family. \\\textit{Scenario 2:} A cluster with $n$ boards from the same FV but with $m$ different families, $m \leq n$. \\\textit{Scenario 3:} A cluster with $n$ boards from various FVs, potentially containing scenarios 1 and 2 within each FV.\\
\autoref{table:scenarios} compares the number of cryptographic operations required by \textit{AgEncID} with \textit{naive} approaches that do not involve cluster formation. A \textit{naive} approach requires individual encryption and decryption for both symmetric bitstream encryption and asymmetric key encryption for each board. \textit{AgEncID} performs best in scenario 1. In scenario 2, the number of \textit{AgEncID} operations increases with more families ($m$), and in scenario 3, \textit{AgEncID} is less efficient, with the number of operations depending on the number of FVs and their board families in a cluster. Our primary focus is on discussing \textit{AgEncID} within scenarios 1 and 2. For example, \autoref{fig:AgEncID-eg} shows how $n = 5$ boards are grouped into two clusters: $Cluster1$ for scenario 1 and $Cluster2$ for scenario 2 within one FV.

\begin{table}[t]
\caption{Possible cluster scenarios for $n$ FPGA boards of $m$ families along with \\the number of operations needed by \emph{AgEncID} vs.\emph{Naive} approaches }  
\centering
\resizebox{\textwidth}{!}{
\begin{tabular}{|c|cccc|cccc|}
\hline
\multirow{3}{*}{\textbf{Operation}} & \multicolumn{4}{c|}{\textbf{Scenario 1}}                                                                                                                                                                                                                                                                                                         & \multicolumn{4}{c|}{\textbf{Scenario 2}}                                                                                                                                                                                                                                                                                                         \\ \cline{2-9} 
                                    & \multicolumn{2}{c|}{Naive Solution}                                                                                                                                      & \multicolumn{2}{c|}{\textbf{AgEncID Solution}}                                                                                                                 & \multicolumn{2}{c|}{Naive Solution}                                                                                                                                      & \multicolumn{2}{c|}{\textbf{AgEncID Solution}}                                                                                                                 \\ \cline{2-9} 
                                    & \multicolumn{1}{c|}{\begin{tabular}[c]{@{}c@{}}Bitstream \\ Encrypt\end{tabular}} & \multicolumn{1}{c|}{\begin{tabular}[c]{@{}c@{}}Key \\ Encrypt\end{tabular}} & \multicolumn{1}{c|}{\begin{tabular}[c]{@{}c@{}}Bitstream \\ Encrypt\end{tabular}} & \begin{tabular}[c]{@{}c@{}}Key \\ Encrypt\end{tabular} & \multicolumn{1}{c|}{\begin{tabular}[c]{@{}c@{}}Bitstream \\ Encrypt\end{tabular}} & \multicolumn{1}{c|}{\begin{tabular}[c]{@{}c@{}}Key \\ Encrypt\end{tabular}} & \multicolumn{1}{c|}{\begin{tabular}[c]{@{}c@{}}Bitstream \\ Encrypt\end{tabular}} & \begin{tabular}[c]{@{}c@{}}Key \\ Encrypt\end{tabular} \\ \hline
Key Generation             & \multicolumn{1}{c|}{$=n$}                                                                     & \multicolumn{1}{c|}{$=n$}                                                               & \multicolumn{1}{c|}{$\mathbf{=1}$}                                                            & $\mathbf{=1}$                                                      & \multicolumn{1}{c|}{$=n$}                                                                     & \multicolumn{1}{c|}{$=n$}                                                               & \multicolumn{1}{c|}{$\mathbf{=1}$}                                                            & $\mathbf{=1}$                                                      \\ \hline
Encryption                 & \multicolumn{1}{c|}{$=n$}                                                                     & \multicolumn{1}{c|}{$=n$}                                                               & \multicolumn{1}{c|}{$\mathbf{=1}$}                                                            & $\mathbf{=1}$                                                      & \multicolumn{1}{c|}{$=n$}                                                                     & \multicolumn{1}{c|}{$=n$}                                                               & \multicolumn{1}{c|}{$\mathbf{\geq m}$, $\mathbf{\leq n}$}                                                                     & $\mathbf{=1}$                                                      \\ \hline
Decryption                 & \multicolumn{1}{c|}{$=n$}                                                                     & \multicolumn{1}{c|}{$=n$}                                                               & \multicolumn{1}{c|}{$=n$}                                                                     & $=n$                                                               & \multicolumn{1}{c|}{$=n$}                                                                     & \multicolumn{1}{c|}{$=n$}                                                               & \multicolumn{1}{c|}{$=n$}                                                                     & $=n$                                                               \\ \hline
\end{tabular}
}
\label{table:scenarios}
\vspace{-0.5cm}
\end{table}

\noindent
\textbf{Initial Setup and Key Generation.} The asymmetric key generation for the $AgEncID$'s key aggregation module are primarily conducted by the FVs.  
First, the FV creates the unique board IDs for each of the boards supplied by her/him, that are hashed to index $i$ such that $i \in {{1,2,3,...,n}}$ and generates the public parameters based on these board IDs. The FV generates a public key $PK$, a master secret key $msk$ for   $AgEncID$  and the private keys $d_i's$ for the $n$ boards. Additionally, FV generates the aggregate key $AgK_{Sj}$  for each cluster. As shown in~\autoref{fig:AgEncID-eg}, the FV receives request for $n$ boards divided into two clusters of boards IDs $j$, $Cluster_1$ with $S_j = \{1,3,4\}$ and $Cluster_2$ with $S_j = \{2,5\}$. While  \textit{Setup()} and \textit{KeyGen()} operations are performed one-time for all $n$ boards, the FV must execute the \textit{Extract()} operation for each cluster separately. Subsequently, the FV transfers the public key $PK$ of all $n$ boards along with the corresponding $AgK_{Sj}$ keys for each cluster to the CSP. In order to enhance security, the FV embeds the private keys $d_i's$ into a tamper-proof non-volatile memory segment of the boards. Finally, each FPGA is equipped with the $AgEncID$ decryption core. 




\begin{algorithm*}[t!]
  \caption{Aggregate Encryption Key Individual Decryption Key --- $AgEncID$  Cryptosystem}
  \begin{algorithmic}[1]
    \Procedure{InitialSetup and KeyGeneration}{FPGA-BOARDS} \Comment{By FVs} 
        \For {\textbf{each} FPGA}
        \State {Assign a unique ID to each of the $n$ FPGA boards} 
        \EndFor
        \State $(param)\gets AgEncID.setup(n)$ \Comment{generates the \emph{AgEncID} cryptosystem parameters}
         \State $(PK,msk)\gets AgEncID.KeyGen(n)$ \Comment{generates the keys for \emph{AgEncID} system}
         \State {Let $N$ denote the set of all IDs for $n$ FPGA boards}   
        \For{\textbf{each} $ID \in N$}
        \State {Generate the board specific private keys $d_i's$}
        \State {Embed $d_i$ in a tamper-proof non-volatile memory segment}
        \EndFor
    \State {Each FPGA is provisioned with a \emph{AgEncID} decryption engine} 
         
    \State {Let $S$ denote the set of all ID-s for a cluster of FPGA boards $S \subseteq N$} 
        \For{\textbf{each} $ID \in S$}
        \State $AgK_{S} \gets AgEncID.Extract(S)$ \Comment{$AgK_S$ is aggregate key for encryption}
        
        \EndFor
    \State {Register the $n$ FPGAs  with CSP in clusters along with their $PK$ and $AgK_{S}$s}  \EndProcedure\\

    \Procedure{Encryption}{Bitstream} \Comment{Performed by IP Providers}
        \State {An IP Provider is assigned with a cluster $S$ of board $IDs$ by CSP}
        \State $K_S \gets AES.KeyGen$ \Comment{$K$ is one-time AES key for $S$ boards}
        \State $EncBit_S \gets AES.Encrypt(Bitstream, K_S)$ \Comment{encrypted bitstream}
        \State $EncK_S = (k_1, k_2, k_3) \gets AgEncID.Encrypt(PK, S, AgK_{S}, K_S)$ \Comment{encrypted key}
        \State {Upload encrypted bitstream $EncBit_S$, ciphertext key triple $EncK_S$ in cloud}
    \EndProcedure\\
    \Procedure{BitstreamDecryption}{Encrypted-Bitstream} 
         \State $K_S \gets AgEncID.Decrypt(S, i, d_i, EncK_S = (k_1, k_2, k_3))$ \Comment{Decrypted AES key}
        \State $bitstream \gets AES.Decrypt(EncBit_S , K_S)$ \Comment{The original bitstream}
    \EndProcedure
  \end{algorithmic}
  \label{algo:AgEncID}
\end{algorithm*}

\smallskip

\noindent
\textbf{Encryption.} The IPP receives requests from the CSP to run IP on a group of boards.  Initially, the IPP creates an encrypted bitstream for the design using its unique AES key specifically for that cluster of boards.~\autoref{fig:AgEncID-eg} shows IPP encrypting the bitstream for $Cluster_j$. Following that, the IPP encrypts the AES key itself using the aggregate key it received for that cluster, $AgK_{Sj}$. The IPP sends both the encrypted bitstream $EncBit$ and the encrypted AES key $EncKey$ to the cloud resource pool. Importantly, this encrypted bitstream can only be decrypted on the FPGA boards within $Cluster_j$.

\smallskip

\noindent
\textbf{Decryption.}
When a cloud user requests a task, the CSP handles the request and forwards the appropriate bitstream to an available FPGA board.
In the decryption process, the board's private key is used for on-board decryption to obtain the AES key, which is then employed within the board's AES decryption core to decrypt the bitstream (depicted in~\autoref{fig:AgEncID-eg} for $Cluster_1$). 
Finally, the FPGA is set up with the decrypted bitstream, and the resulting output is delivered to the user.
\section{Analysis of $AgEncID$ Security and Performance}
\label{sec:analysis}
In this section, we delve into the security and performance aspects of the \textit{AgEncID} framework.

\subsection{$AgEncID$ Security}
The proposed \textit{AgEncID} offers security against the threat model described in Section~\ref{sec:threat-model}. The security of $AgEncID$ relies on the security of the AES-256 cipher and the key aggregation scheme. AES-256 is a strong encryption algorithm that has been subjected to extensive cryptanalysis and no significant weaknesses have been found. AES-256 key is protected by $AgEncID$ with an elliptic curve (EC)-based construction that is provably secure against chosen plaintext attacks (CPAs) on the key \cite{kac} under the discrete logarithm problem (DLP) in EC and the bilinear Diffie-Hellman exponent (BDHE) problem \cite{ecccrypto}. The public and secret key parameters for the \emph{AgEncID} system are protected under the security assumptions of DLP in EC and BDHE \cite{collusion}.

\smallskip

\noindent
\textbf{Security of Bitstream.} The bitstream is encrypted with AES-256 by IPP using its own key and sent to the CSP over secure channels such as SSL or TLS to prevent interception through \emph{MiM} attacks. The AES-256 encryption provided by major FPGA vendors is resistant to conventional attacks such as \emph{cloning}, \emph{RE}, unauthorized copying of the bitstream, and \emph{tampering}. Decryption of the bitstream is exclusively possible on the FPGA chip, using the AES key known only to its owner. This AES key is further encrypted with \textit{AgEncID}, ensuring that it can only be decrypted on-chip using FPGA's embedded private key. Hence, neither the CSP nor a malicious external entity can access the unencrypted bitstream. Even if IPPs and CUs were to collude in an attempt to expose other authorized IPPs or CUs' bitstreams, their efforts would be futile. Accessing the AES key is only feasible on the FPGA chip itself, and the storage and operations on these FPGAs are considered highly secure. The vendors claim that their FPGAs can defend against attacks that occur while the FPGA is operating, such as \emph{SCA} and the leakage of protected information across internal boundaries. 

\smallskip

\noindent 
\textbf{Security of Key.}
In the \textit{AgEncID} framework, the AES key provisioning to the FPGA is highly secure, thwarting unauthorized interception. The \textit{AgEncID} scheme's CPA-secure nature makes it exceptionally resilient, even if the encrypted key is compromised. This key is stored securely alongside the bitstream in the cloud, albeit vulnerable to potential \textit{MiM} attacks during transit.
Decryption relies on the device-specific private key inside the secure FPGA storage. This key is inaccessible to everyone, including the CSP and IPPs. Additionally, the key provisioning process does not involve TTPs, which reduces the risk of \textit{MiM} attacks.
AES key decryption occurs on the FPGA through \textit{AgEncID}, with the decrypted key briefly stored in \text{Battery-Backed RAM} (BBRAM), reliant on continuous power for data retention. Tampering attempts lead to power cuts, erasing \text{BBRAM} and the FPGA configuration, preventing \text{Side-Channel Attacks} (SCA) and tampering\cite{aes}. Additionally, Xilinx (AMD) ensures that the key cannot be read or reprogrammed externally once programmed, limiting access solely to the internal bitstream decryption engine (\url{www.docs.xilinx.com/v/u/en-US/xapp1239-fpga-bitstream-encryption}). Cloud-FPGA communication employs secure protocols, granting CSP logical access for processing without hardware access. IPs are provided to users as an Application-as-a-Service, eliminating external network connections and mitigating side-channel attack risks. While this work doesn't address high-level adversaries targeting FPGA fabrics, it establishes robust security. \textit{AgEncID} guarantees that the AES key remains inaccessible to any unauthorized individuals, including CUs and external agents.

\subsection{AgEncID Performance and Efficiency}
The AES algorithm has the same computational complexity as naive AES-256. We discuss the algorithmic efficiency of the key aggregation method of \textit{AgEncID} below.

\smallskip

\noindent
\textbf{Low overhead Key Generation.} 
$AgEncID.setup()$ operation has a cost of $\mathcal{O}(|n|)$ for generating the cryptosystem parameters of size $\mathcal{O}(|n|)$. The size of the public key, which depends on these parameters, scales linearly with the total number of FPGA boards. This is generally manageable in organizations with ample shared storage. Each board is assigned a private key consisting of just one group element, resulting in low and constant-size private keys.
A single AES key is used to encrypt bitstreams for a cluster of $|S| \leq n$ boards. Extracting the fixed size aggregate key for encrypting AES keys for $|S| \leq n$ boards requires $\mathcal{O}(|S|)$ group additions and is of fixed size. Consequently, \textit{AgEncID}'s key generation is both time and space-efficient.

\smallskip

\noindent
\textbf{Efficient Encryption and Decryption.} Encryption doesn't require pairing operations, the pre-computed value $\hat{e}(g_n,tg_1)$ can be quickly processed at a constant rate. The resulting ciphertext is compact in size. Decryption, on the other hand, involves $\mathcal{O}(|S|)$ additions in the group and two pairing operations.  
Recent high-speed software implementations (\url{www.crypto.stanford.edu/pbc}, \url{www.github.com/herumi/mcl}, etc.) enable fast bilinear pairing computations, even when employed on devices with limited computational resources.
Furthermore, both the encryption and decryption routines in AgEncID require a fixed number of elliptical curve operations.

\smallskip

\noindent
\textbf{Flexible Resource Management by CSP.}
The CSPs should have the flexiblity in provisioning bitstreams and keys to FPGAs. $AgEncID$ encryption scheme allows IPPs to encrypt their bitstreams and their keys for a set of FPGAs as requested by CSP. Thus CSPs can manage the FPGAs and assign different FPGAs to users at each run in the cloud according to resource availability.
\vspace{-0.2cm}

\subsection{Discussion}
The CSP may wish to add more FPGAs in the cloud to satisfy users' demands. In such a scenario, the CSP would contact the FV for a new board registration with the cloud. The FPGA vendor then manages the $AgEncID$ parameters for a new board. Dynamic board registration at the FPGA vendor side is managed through the unique board IDs $i$ that belongs to a set $S$.  The FV maintains a flag to monitor availability of previously assigned $i$ in case of de-registration of a board from the set $S$. Available board IDs $i$'s are assigned to the new boards dynamically during board registration. In such a scenario, the public key $PK$ does not need to be updated. With a new board registrations under new ID, the FV expands the public key $PK$ by simply adding two more elements to the $param$ of $PK$. The secret parameter $\alpha$ during setup is randomly generated. $\alpha$ is made event driven and generated whenever a new board is registered. Thus the CSP need not contact the IP developer to use their IP on the newly added FPGAs unless a new board ID is added in $S$.

\section{Experiments and Evaluation}
\label{sec:experiments}

\subsection{Experimental Setup}
Experiments are performed on 64-bit linux machine, having Intel Core i5 processor and 16 GB RAM clocked at 3.2 GHz.  For our evaluation, we utilized five Xilinx boards from various families, as indicated in ~\autoref{tab:boards}. Vivado Design Suite is used to generate bitstream from a HDL design and perform simulations. 
A pairing friendly elliptic curve construction is needed for the bilinear group mapping. We have used Type-A curves bundled with Pairing-Based Cryptography, PBC version 0.5.14 (\url{www.crypto.stanford.edu/pbc}). Four properties of the curve used for symmetric bilinear pairing are: (\textit{i}) Type A supersingular curve \(y^2 = x^3 + x\); (\textit{ii}) $160$-bit Solinas prime number \emph{p}, which offers $1024$-bit of discrete-logarithm security; (\textit{iii}) the embedding degree $k$ of pairing is $2$; (\textit{iv}) elements of group $G$ and $G_T$ take $512$ and $1024$ bits respectively.

\begin{table}[t]
\centering
\caption{Properties of FPGA boards used in Experiments} 
\resizebox{\textwidth}{!}{
\begin{tabular}{|c|c|c|c|c|c|}
\hline
\textbf{No.} & \textbf{Board} & \textbf{Family}                                                         & \textbf{Device}   & \textbf{Processor}                                                                        & \textbf{Memory}                                                        \\ \hline
B1           & ZC702          & \multirow{3}{*}{\begin{tabular}[c]{@{}c@{}}F1\\ Zynq-7000\end{tabular}} & XC7Z020-CLG484-1  & Dual ARM Cortex-A9 core                                                                   & 1GB DDR3                                                               \\ \cline{1-2} \cline{4-6} 
B2           & PYNQ-Z2        &                                                                         & XC7Z020-1CLG400C  & Dual ARM® Cortex™-A9 MPCore™                                                              & \begin{tabular}[c]{@{}c@{}}256 KB on-chip,\\ 8 GB SD card\end{tabular} \\ \cline{1-2} \cline{4-6} 
B3           & Zybo-Z7        &                                                                         & XC7Z020-1CLG400C  & Dual-core ARM Cortex-A9                                                                   & 1 GB DDR3L                                                             \\ \hline
B4           & ZCU104         & \begin{tabular}[c]{@{}c@{}}F2\\ Zynq UltraScale+ MPSoC\end{tabular}     & XCZU7EV-2FFVC1156 & \begin{tabular}[c]{@{}c@{}}quad-core ARM® Cortex™-A53,\\ Dual-core Cortex-R5\end{tabular} & PS DDR4 2GB                                                            \\ \hline
B5           & KC705          & \begin{tabular}[c]{@{}c@{}}F3\\ Kintex 7\end{tabular}                   & XC7K325T-2FFG900C & MicroBlaze, 32bit RISC                                                                    & 1GB DDR3 SODIM                                                         \\ \hline                                                        
\end{tabular}
}
\label{tab:boards}
\vspace{-0.2cm}
\end{table}

\subsection{Implementation and tools used} Our scheme relies on two cryptographic modules: the standard \emph{AES} cores offered by FPGA vendors and the key aggregation cores of \emph{AgEncID}. In this section, we'll describe how we implemented and utilized these modules within the \emph{AgEncID} framework, including the tools we employed.

\smallskip

\noindent
\textbf{Key Generation and Encryption.}
To generate the 256-bit AES encryption key for encrypting a bitstream, standard tools and techniques can be employed. In our case, we used Vivado to create the specific AES key provided by the vendor. Bitstream encryption was carried out using the default AES-256 encryption feature available in Xilinx Vivado. 
To encrypt the 256-bit AES key, we utilized $AgEncID$'s key aggregation encryption module. This module is implemented in software using the GNU multiple precision arithmetic library (GMP) (\url{www.gmplib.org}, version 6.2.1) and the PBC Library (\url{www.crypto.stanford.edu/pbc}). PBC Library is used for the underlying elliptic-curve group and pairing operations. PBC provides the necessary APIs to compute pairings over the BN family of elliptical curves. The $\mathbf{Encrypt(PK, S, K_S, m)}$ operation of key aggregation module in \emph{AgEncID} is implemented through three primary operations: point addition, point doubling and pairing with CPU latencies of $3.01 \times 10^{-3}$ ms, $6.2 \times 10^{-3}$ ms and $19.4$ ms respectively.

\begin{figure}[t]
    \centering
    \resizebox{0.75\textwidth}{!}{\includegraphics[]{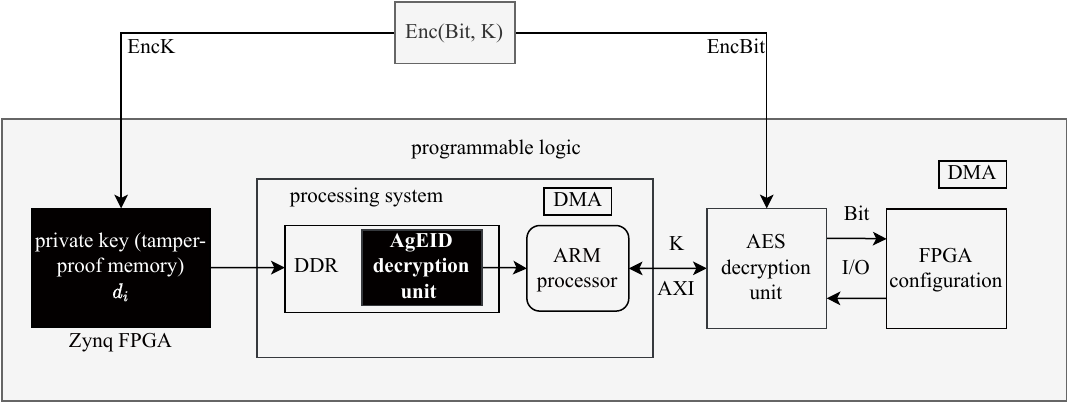}}
    \caption{Flowchart for $AgEncID$'s $Decrypt()$ operation in key-aggregation module}
    \label{fig:ps}
    \vspace{-0.2cm}
\end{figure}

\smallskip

\noindent
\textbf{Decryption.}
Similar to the encryption phase, the decryption process in \textit{AgEncID} also consists of two parts: key decryption and bitstream decryption.
For key decryption, the $\mathbf{Decrypt(S, i, d_i, C = (c_1, c_2, c_3))}$ operation of $AgEncID$'s key aggregation module involves elliptic curve pairing operations. 
For the decryption operation we utilize an embedded processor core in the Processing System (PS) section (eg. B1-B4 in~\autoref{tab:boards}) of modern SoCs.
This decryption module leverages the same libraries (GMP and PBC) used for $AgEncID$'s $Encrypt()$ operation, and is deployed in the PS region of the FPGAs (as shown in~\autoref{fig:ps}). 
On $ARM Cortex-A9$ processors from the $Zynq-7000$ family, this software module takes  $401$ ms to decrypty an AES-256 key.
For SoCs lacking an embedded processor core in the PS, we've developed a hardware core for $AgEncID$'s $Decrypt()$ operation written in Verilog. 
This core can be deployed in the Programmable Logic (PL) of the FPGA. It implements a pairing-friendly elliptic curve providing a 256-bit security level~\cite{pairingecc}, including point addition and scalar multiplication operations.
The bilinear pairing core used is inspired by the Duursma-Lee algorithm~\cite{lee} for pairing operations.
For bitstream decryption, we utilize the on-chip AES decryption module provided by vendors. This module has standard resource requirements---Xilinx AES decryption core (\url{www.xilinx.com/htmldocs/ip_docs/pru_files/aes.html}).

\subsection{Results and Evaluation}
We evaluate the effectiveness of the \emph{AgEncID} cryptosystem in three key aspects: (\textit{i}) execution time, (\textit{ii}) energy consumption in terms of CPU usage, and (\textit{iii}) FPGA resource overhead. 
To gauge its performance, we compared it with the one-to-one mechanism used in existing approaches, referred as \textit{IEID} (Individual Encryption Individual Decryption). 
Our evaluation for bitstream encryption was conducted using a set of experiments on bitstreams generated for the \textit{ISCAS} (\url{www.sportlab.usc.edu/~msabrishami/benchmarks.html}) benchmark circuits 
On the other hand, key encryption is performed on 256-bit AES key.
We conduct following experiments:

\begin{itemize}
    \item \textit{Experiment 1.} We conducted bitstream encryption for a cluster of $20$ boards, representing scenario 1, using \emph{AgEncID} and \emph{IEID}. We used the $C17$ benchmark design and selected the $20$ boards from the F1 family (specifically, boards B1, B2, and B3), as listed in~\autoref{tab:boards}.
    \item \textit{Experiment 2.} Bitstream encryption was carried out on $C17$ for scenario 2. A cluster of $10$ boards are chosen from three different families F1, F2, and F3 as listed in \autoref{tab:boards}, i.e $n = 10$ and $m = 3$. We take  $3$ boards from F1 and F2 each and $4$ boards from F3. We again compared the \textit{AgEncID} approach with the \textit{IEID} approach.
    \item \textit{Experiment 3.} We performed bitstream encryption for a cluster of $5$ boards, depicting scenario 1, using both the \emph{AgEncID} and \emph{IEID} approaches. We evaluated on three benchmark designs of varying sizes: $C432$, $C499$, and $C880$.
    \item \textit{Experiment 4.} We performed 256-bit AES key encryption using key aggregation module of AgEncID for $20$ boards, covering scenarios 1 and 2. Additionally, we performed 256-bit AES key encryption by altering key aggregation module of \emph{AgEncID} according to \emph{IEID} framework.
\end{itemize}

\begin{figure}[]
\centering





    \subfloat[\tiny AES Encryption: scenario 1]{
    \begin{tikzpicture}[scale=0.3] 
\begin{axis}
[    
height=0.8\textwidth,
width=1\textwidth,
xtick = {0,5,10,15,20, 25},
xticklabels = {0,5,10,15,20, 25},
ytick = {0,5,10,15,20, 25, 30},
yticklabels = {0,5,10,15,20, 25, 30},
axis line style = very thick,
tick label style={},
label style={},
xlabel=\textbf{Number of Boards},ylabel=\textbf{Execution Time (s)},
grid style = dashed,
grid=both,
legend style=
{at={(0.5, .9)}, 
anchor=north east,
} ,
]

\addplot[color=black, solid, mark=FPGA, line width=2] plot coordinates {

(2, 1.19)
(4, 1.19)  
(8, 1.19) 
(10, 1.19)
(20, 1.19) 
};



\addplot[color=blue, solid, mark=FPGA, line width=2] plot coordinates {

(2, 2.38)
(4, 4.76)  
(8, 9.52) 
(10, 11.9)
(20, 23.8) 
};

\legend{\huge{\bf{AgEncID}}\\\huge{\bf{IEID}}\\}
\end{axis}
\end{tikzpicture}
}
\subfloat[\tiny AES Encryption: scenario 2]{
    \begin{tikzpicture}[scale=0.3] 
    \begin{axis}
    [    
height=0.8\columnwidth,
width=1\columnwidth,
axis line style = very thick,
tick label style={},
xtick = {0,2,4,6,8,10,12},
xticklabels = {0,2,4,6,8,10,12},
ytick = {0,5,10,20,30},
yticklabels = {0,5,10,20,30},
label style={},
xlabel=\textbf{Number of Boards},ylabel=\textbf{Execution Time (s)},
grid style = dashed,
grid=both,
legend style=
{at={(0.5, 0.9)}, 
anchor=north east,
} ,
]

\addplot[color=black, solid, mark=FPGA, line width=2] plot coordinates {

(2, 1.19) 
(4, 5.2) 
(6, 5.2) 
(8, 8.44)
(10, 8.44)
};

\addplot[color=blue, solid, mark=FPGA, line width=2] plot coordinates {

(2, 2.38) 
(4, 7.56) 
(6, 15.58) 
(8, 22.06)
(10, 28.56)
};

\legend{\huge{\bf{AgEncID}}\\\huge{\bf{IEID}}\\}
\end{axis}
\end{tikzpicture}
}
\subfloat[\tiny Time vs bitstream size: scenario 1]{
\pgfplotstableread[row sep=\\,col sep=&]{
    interval & column1 & column2\\
    C432 48     & 2.52  & 12.6\\
    C499 60     & 4.204 & 21.02 \\
    C880 111    & 9.285 & 46.4 \\
    }\time

\begin{tikzpicture}[scale=0.3]
    \begin{axis}[
            ybar,
            axis line style = very thick,
            tick label style={align=center}, 
            label style={}, 
            bar width=1cm,
            width=\columnwidth,
            height=.8\columnwidth,
            legend style={at={(0.5, .9)}, anchor=north east},
            symbolic x coords={C432 48, C499 60, C880 111},
            xticklabels={C432(48MB), C499 (60MB), C880(111MB)},
            xtick=data,
            ymin=0,ymax=50,
            ylabel={\textbf{Execution Time (s)}},
            xlabel={\textbf{Design bitstream with size}},
        ]
        \addplot table[x=interval,y=column1]{\time};
        \addplot table[x=interval,y=column2]{\time};
        \legend{\textbf{\huge AgEncID}, \textbf{\huge IEID}}
    \end{axis}
\end{tikzpicture}

}
 \caption{Execution time of \textit{AgEncID} and \textit{IEID} for bitstream encryption}
 \label{fig:aes-energy-time}
\end{figure}
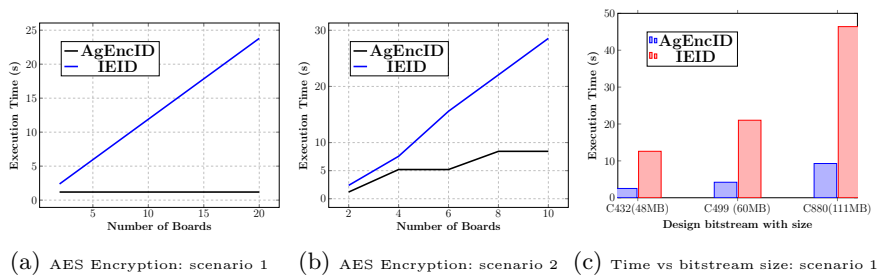

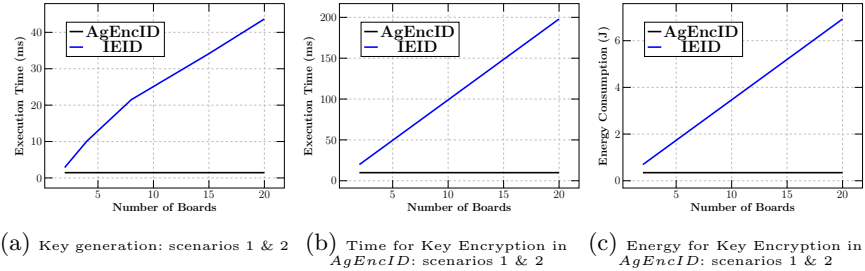
\begin{figure*}[h]
\centering
\subfloat[\tiny Key generation: scenarios 1 \& 2]{
    \begin{tikzpicture}[scale=0.3] 
\begin{axis}
[    
height=0.8\columnwidth,
width=1\columnwidth,
every major tick/.append style={very thick, major tick length=10pt, black},
axis line style = very thick,
tick label style={},
xtick = {0,5,10,15,20, 25},
xticklabels = {0,5,10,15,20, 25},
ytick = {0,10,20,30,40},
yticklabels = {0,10,20,30,40},
label style={},
xlabel=\textbf{Number of Boards},ylabel=\textbf{Execution Time (ms)},
grid style = dashed,
grid=both,
legend style=
{at={(0.5, .9)}, 
anchor=north east,
} ,
]

\addplot[color=black, solid, mark=FPGA, line width=2] plot coordinates {

(2, 1.45)
(4, 1.45)  
(8, 1.45) 
(15, 1.45)
(20, 1.45) 
};

\addplot[color=blue, solid, mark=FPGA, line width=2] plot coordinates {

(2, 2.87)
(4, 10.12)  
(8, 21.49) 
(15, 34.14)
(20, 43.67) 
};

\legend{\huge{\bf{AgEncID}}\\\huge{\bf{IEID}}\\}
\end{axis}
\end{tikzpicture}
}
\subfloat[\tiny Time for Key Encryption in $AgEncID$: scenarios 1 \& 2]{
    \begin{tikzpicture}[scale=0.3] 
\begin{axis}
[    
height=0.8\columnwidth,
width=1\columnwidth,
every major tick/.append style={very thick, major tick length=10pt, black},
axis line style = very thick,
tick label style={},
xtick = {0,5,10,15,20, 25},
xticklabels = {0,5,10,15,20, 25},
ytick = {0,50,100,150,200, 250},
yticklabels = {0,50,100,150,200, 250},
label style={},
xlabel=\textbf{Number of Boards},ylabel=\textbf{Execution Time (ms)},
grid style = dashed,
grid=both,
legend style=
{at={(.5,.9)}, 
anchor=north east,
anchor= north east ,
} ,
]

\addplot[color=black, solid, mark=FPGA, line width=2] plot coordinates {

(2, 9.9)
(4, 9.9)  
(8, 9.9) 
(10, 9.9)
(20,9.9) 
};

\addplot[color=blue, solid, mark=FPGA, line width=2] plot coordinates {

(2, 19.8)
(4, 39.6)  
(8, 79.2) 
(10, 99)
(20, 198) 
};

\legend{\huge{\bf{AgEncID}}\\\huge{\bf{IEID}}\\}
\end{axis}
\end{tikzpicture}
}
\subfloat[\tiny Energy for Key Encryption in $AgEncID$: scenarios 1 \& 2]{
    \begin{tikzpicture}[scale=0.3] 
\begin{axis}
[    
height=0.8\columnwidth,
width=1\columnwidth,
every major tick/.append style={very thick, major tick length=10pt, black},
axis line style = very thick,
tick label style={},
xtick = {0,5,10,15,20, 25},
xticklabels = {0,5,10,15,20, 25},
ytick = {0,2, 4, 6, 8, 10},
yticklabels = {0,2, 4, 6, 8, 10},
label style={},
xlabel=\textbf{Number of Boards},ylabel=\textbf{Energy Consumption (J)},
grid style = dashed,
grid=both,
legend style=
{at={(.5,0.9)}, 
anchor=north east,
anchor= north east ,
} ,
]

\addplot[color=black, solid, mark=FPGA, line width=2] plot coordinates {

(2, 0.3465)
(4, 0.3465)  
(8, 0.3465) 
(10, 0.3465)
(20, 0.3465) 
};

\addplot[color=blue, solid, mark=FPGA, line width=2] plot coordinates {

(2, 0.693)
(4, 1.386)  
(8, 2.772) 
(10, 3.465)
(20, 6.930) 
};

\legend{\huge{\bf{AgEncID}}\\\huge{\bf{IEID}}\\}
\end{axis}
\end{tikzpicture}
}

 \caption{\textit{AgEncID} vs \textit{IEID} in terms of key generation and aggregation}
 \label{fig:AgEncID-energy-time}
 \vspace{-0.5cm}
\end{figure*}

\noindent \textbf{Execution Time.} 
In~\autoref{fig:aes-energy-time}, we depict the CPU execution time for \textit{AgEncID}'s AES bitstream encryption. In~\autoref{fig:aes-energy-time} (a), we observe that \textit{AgEncID} consistently requires the same amount of time in \textit{Experiment 1}, in contrast to the \textit{IEID} approach, where the time increases linearly as the number of boards increases.~\autoref{fig:aes-energy-time} (b) depicts that \textit{AgEncID}'s time requirement varied for AES encryption within the different board families (\textit{Experiment 2}). 
Despite the differing time requirements, the \textit{AgEncID} approach maintains its superiority over the \textit{IEID} approach.
In \textit{Experiment 2}, \textit{AgEncID}'s performance is similar to \textit{IEID} for bitstream encryption when $n = m$. 
~\autoref{fig:aes-energy-time} (c) depicts a significant $5X$ time difference between \textit{IEID} and \textit{AgEncID} in \textit{Experiment 3}, which represents a substantial cost when handling large bitstreams. These results are derived from Vivado CPU timing report. 
For both scenarios 1 and 2, \textit{AgEncID} maintains a constant time requirement in \textit{Experiment 4}, unlike \textit{IEID} that requires individual key encryption for each board.
\autoref{fig:AgEncID-energy-time} (a) illustrates that symmetric key generation is performed at a constant time by \textit{AgEncID} for any number of boards, in contrast to \textit{IEID} approach.
\autoref{fig:AgEncID-energy-time} (b) provides timing results for \textit{Experiment 4}, indicating the constant CPU time needed for executing \textit{AgEncID}'s key encryption algorithm unlike \textit{IEID} that requires individual key encryption for each board. 

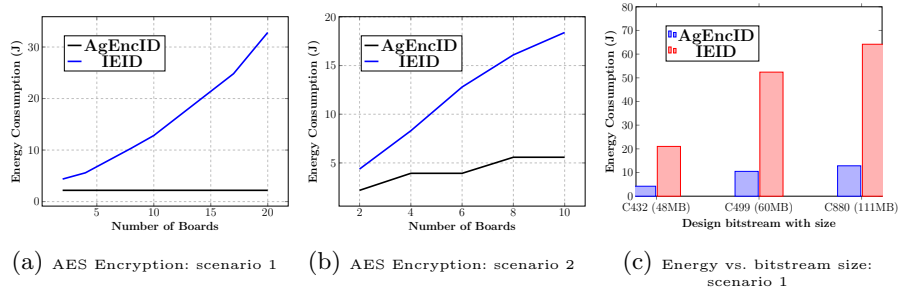
\begin{figure}[]
\centering
\resizebox{\textwidth}{!}{
\subfloat[\tiny AES Encryption: scenario 1]{
    \begin{tikzpicture}[scale=0.3] 
    \begin{axis}
    [    
    height=0.8\columnwidth,
    width=1\columnwidth,
    axis line style = very thick,
    tick label style={},
    xtick = {0,5,10,15,20, 25},
    xticklabels = {0,5,10,15,20, 25},
    ytick = {0,10,20,30,40},
    yticklabels = {0,10,20,30,40},
    label style={},
    xlabel=\textbf{Number of Boards},
    ylabel=\textbf{Energy Consumption (J)},
    grid style = dashed,
    grid=both,
    legend style=
    {at={(0.5, 0.9)}, 
    anchor=north east,
    } ,
    ]

    \addplot[color=black, solid, mark=FPGA, line width=2] plot coordinates {
    
    (2, 2.185) 
    (4, 2.185) 
    (8, 2.185) 
    (10, 2.185)
    (20, 2.185)
    };

    \addplot[color=blue, solid, mark=FPGA, line width=2] plot coordinates {
    
    (2, 4.37) 
    (4, 5.586) 
    (8, 10.326) 
    (10, 12.82)
    (17, 24.82)
    (20, 32.82)
    };

    \legend{\huge{\bf{AgEncID}}\\\huge{\bf{IEID}}\\}
    \end{axis}
    \end{tikzpicture}
}
\subfloat[\tiny AES Encryption: scenario 2]{
    \begin{tikzpicture}[scale=0.3] 
    \begin{axis}
    [    
height=0.8\columnwidth,
width=1\columnwidth,
axis line style = very thick,
tick label style={},
xtick = {0,2,4,6,8,10,12},
xticklabels = {0,2,4,6,8,10,12},
ytick = {0,5,10, 15, 20},
yticklabels = {0,5,10, 15, 20},
label style={},
xlabel=\textbf{Number of Boards},ylabel=\textbf{Energy Consumption (J)},
grid style = dashed,
grid=both,
legend style=
{at={(0.5, 0.9)}, 
anchor=north east,
} ,
]

\addplot[color=black, solid, mark=FPGA, line width=2] plot coordinates {

(2, 2.185) 
(4, 3.932) 
(6, 3.932) 
(8, 5.582)
(10, 5.582)
};

\addplot[color=blue, solid, mark=FPGA, line width=2] plot coordinates {

(2, 4.37) 
(4, 8.302) 
(6, 12.796) 
(8, 16.096)
(10, 18.396)
};

\legend{\huge{\bf{AgEncID}}\\\huge{\bf{IEID}}\\}
\end{axis}
\end{tikzpicture}
}
\subfloat[\tiny Energy vs. bitstream size: scenario 1]{

    \pgfplotstableread[row sep=\\,col sep=&]{
    interval & column1 & column2\\
    C432(48MB)     & 4.204  & 21.02\\
    C499(60MB)    & 10.48 & 52.4 \\
    C880(111MB)   & 12.835 & 64.175 \\
    }\energy

    \begin{tikzpicture}[scale=0.3]
    \begin{axis}[
            height=0.8\columnwidth,
            width=1\columnwidth,
            ybar,
            axis line style = very thick,
            tick label style={align=center},
            label style={},
            bar width=1cm,
            legend style={at={(0.5, 0.9)}, anchor=north east},
            symbolic x coords={C432(48MB), C499(60MB), C880(111MB)},
            xticklabels={C432 (48MB), C499 (60MB), C880 (111MB)},
            xtick=data,
            ymin=0,ymax=80,
            ylabel={\textbf{Energy Consumption (J)}},
            xlabel={\textbf{Design bitstream with size}},
        ]
        \addplot table[x=interval,y=column1]{\energy};
        \addplot table[x=interval,y=column2]{\energy};
        \legend{\textbf{\huge AgEncID}, \textbf{\huge IEID}}
    \end{axis}
\end{tikzpicture}
}}

\caption{Energy consumption by \textit{AgEncID} and \textit{IEID} for bitstream encryption}
\label{fig:AgEncID-energy}
\end{figure}
\smallskip
\noindent \textbf{Energy Consumption.} \autoref{fig:AgEncID-energy} illustrates the comparison between \emph{AgEncID} and \textit{IEID} approaches in terms of energy consumption for \textit{Experiment 1}, similar to the comparison of their respective execution times. 
From \autoref{fig:AgEncID-energy} (a), it is evident that \emph{AgEncID} consistently demonstrates superior energy efficiency over \emph{IEID}, regardless of the number of boards in the cluster. 
Furthermore, \autoref{fig:AgEncID-energy} (c) highlights the energy requirements for \emph{AgEncID} and \emph{IEID} in \textit{Experiment 3}, showcasing substantial energy savings achieved by \emph{AgEncID}. 
These energy calculations are obtained from the Vivado CPU power consumption report. 
The results for \textit{Experiment 2} and \textit{Experiment 4} for energy consumption is similar to that of execution time which is depicted in ~\autoref{fig:AgEncID-energy} (b) and \autoref{fig:AgEncID-energy-time}(c) respectively. 
Similar to execution time, when it comes to energy consumption, key encryption demonstrates significantly lower resource utilization compared to bitstream encryption.

\smallskip

\noindent \textbf{FPGA Resource Overhead.}
\textit{AgEncID}'s key aggregation algorithm primarily operates in external software, reducing FPGA resource demands. 
Encryption tasks are conducted externally to the FPGA.
The FPGA only loads the symmetric AES key when necessary. 
A few kilobytes (KBs) is required for the board-specific private key and \textit{AgEncID}'s key decryption module running in the processor core of PS. 
Depending on device availability and licensing, we've deployed the \textit{AgEncID}'s key decryption module on PL of board $B1$ (~\autoref{tab:boards}). 
In~\autoref{table:resource-util}, an overview of FPGA's resource utilization, latency and power consumption of \textit{AgEncID} hardware decryption module is reported.
Directly comparing FPGA resource overhead of \textit{AgEncID} with existing techniques is challenging due to differences in evaluation environments and FPGA features. 
However, in terms of parameters such as LUTs, Registers, and Slices, $AgEncID$'s resource overhead typically falls in the range of $2$\% to $17$\%, whereas related works have resource overhead ranging from $2$\% to $51$\%~\cite{fpga-trust,saar,multitenantFpga}.

\begin{table}[t]
\centering
\small
\caption{Resource utilization, storage, performance and power consumption report for \emph{AgEncID}'s $Decrypt()$ operation of key-aggregation, on \textit{ZC702}} 
\resizebox{0.9\textwidth}{!}{
\begin{tabular}{@{}cccccccc@{}}
\toprule
\textbf{Module} & \multicolumn{3}{c}{\textbf{Resource Utilization}} & \textbf{Storage (KB)} & \multicolumn{2}{c}{\textbf{Performance}} & \textbf{Power (Watt)}\\
\cmidrule(lr){2-4} \cmidrule(lr){6-7}
& \begin{tabular}[c]{@{}l@{}} \textbf{LUTs} \end{tabular} &  \begin{tabular}[c]{@{}l@{}} \textbf{Registers} \end{tabular} & \begin{tabular}[c]{@{}l@{}} \textbf{Slices} \end{tabular} & & \begin{tabular}[c]{@{}l@{}} \textbf{\# Clock Cycles} \end{tabular} & \begin{tabular}[c]{@{}l@{}} \textbf{Latency (ns)}  \end{tabular} & \\
\midrule
\multirow{2}{*}{ECC} & $1181$ (\#) & $1403$ (\#) & $503$ (\#) & \multirow{2}{*}{$7695.402$} & \multirow{2}{*}{$2415$} & \multirow{2}{*}{$6.667$} & \multirow{2}{*}{$0.114$} \\
& $2.21$ (\%) & $1.32$ (\%) & $3.78$ (\%) & & & & \\
\midrule
\multirow{2}{*}{Pairing} & $7608$ (\#) & $13401$ (\#) & $1668$ (\#) & \multirow{2}{*}{$8119.027$} & \multirow{2}{*}{$57456$} & \multirow{2}{*}{$20$} & \multirow{2}{*}{$0.192$} \\
& $14.35$ (\%) & $12.59$ (\%) & $12.54$ (\%) & & & & \\
\bottomrule
\end{tabular}
}
\label{table:resource-util}
\end{table}

\section{Conclusion}
\label{sec:conclude}
This paper present \textit{AgEncID} (\textit{Ag}gregate \textit{E}ncryption and \textit{I}ndividual \textit{D}ecryption) cryptosystem for FPGA bitstream protection through secure key provisioning.
\textit{AgEncID} employs a new key aggregation technique and a streamlined encryption-decryption process to significantly reduce the resource overhead (storage, time, and energy) of both key and bitstream encryption, compared to existing methods.
In the future, we plan to implement cutting-edge cryptographic algorithms to address threats arising from advanced on-board FPGA attacks and large-scale quantum computing attacks.

\bibliographystyle{splncs04}
\bibliography{mybibliography}
\end{document}